\documentclass[letterpaper,twocolumn,10pt]{article}
\usepackage{times}
\usepackage{balance}
\usepackage{multirow}
\usepackage{graphics}
\usepackage{graphicx}
\usepackage{cite}
\usepackage{textcomp}
\usepackage{listings}
\usepackage[hyphens]{url}
\usepackage{hyperref}
\hypersetup{
 colorlinks=false,
 pdfborder={0 0 0},
}
\usepackage{todonotes}
\usepackage{authblk}
\usepackage{subcaption}
\usepackage{epstopdf}
\usepackage{flushend}
\usepackage{mdwlist}

\begin{document}

\title{Master of Puppets:\\ Analyzing And Attacking A Botnet For Fun And Profit}
\author{Genki Saito and Gianluca Stringhini}
\affil{University College London\\ \{genki.saito.12, g.stringhini\}@ucl.ac.uk}
\date{}

\maketitle

\begin{abstract}
  A botnet is a network of compromised machines (bots), 
under the control of an attacker. Many of these machines are infected 
without their owners' knowledge, and botnets are the driving force behind 
several misuses and criminal activities on the Internet (for example spam emails). 
Depending on its topology, a botnet can have zero or more command and control (C\&C) servers, 
which are centralized machines controlled by the cybercriminal that issue 
commands and receive reports back from the co-opted bots.

In this paper, we present a comprehensive analysis of the command and control 
infrastructure of one of the world's largest proprietary spamming botnets 
between 2007 and 2012: Cutwail/Pushdo. 
We identify the key functionalities needed by a spamming botnet to operate effectively. 
We then develop a number of attacks 
against the command and control logic of Cutwail that target those functionalities, and make the spamming operations of the botnet
less effective. 
This analysis was made possible by having access to the source code 
of the C\&C software, as well as setting up our own Cutwail C\&C server, and by 
implementing a clone of the Cutwail bot.
With the help of this tool, we were able to enumerate the number of bots currently 
registered with the C\&C server, impersonate an existing bot to report false 
information to the C\&C server, and manipulate spamming statistics of an arbitrary bot
stored in the C\&C database.
Furthermore, we were able to make the control server inaccessible by conducting a 
distributed denial of service (DDoS) attack.
Our results may be used by law enforcement and practitioners to develop better
techniques to mitigate and cripple other botnets, since many of findings are
generic and are due to the workflow of C\&C communication in general.
\end{abstract}

\section{Introduction}

Botnets, networks of compromised computers under the control of the same
cybercriminal, have been the tool of choice of miscreants committing illicit
actions on the Internet for the last 10
years~\cite{Cooke:2005:ZRU,nadji2013beheading,rossow2013sok}.
Security researchers and law enforcement experts are constantly engaged in an
arms race with cybercriminals, finalized to disrupt botnet
operations~\cite{nadji2013beheading,StoneGross:11:leet}.
Unfortunately, this arms race is difficult to win, because cybercriminals
have the advantage that they can react to the countermeasures deployed by the
security community and make their botnets more resilient to
takedowns~\cite{stone:09:torpig}. Moreover, the fact that botnet operations
are distributed across the globe, and that different critical parts of the
malicious infrastructure are typically located in different countries makes it
particularly difficult for law enforcement to effectively coordinate and take
down such operations~\cite{stringhini2014harvester,liu2011effects}.

Due to the complexity of the botnet phenomenon, a wealth of research has been
conducted on understanding such cybercriminal operations. A category of work
focuses on understanding the monetization of
botnet
operations~\cite{Caballero2011,kanich:08:spamalytics,Kanich2011,levchenko2011click}.
Botnets need to generate a profit for their administrator (botmaster), and this
usually happens by renting them out to other cybercriminals or by using them
directly to perform illicit activities such as sending email spam or stealing
financial information from the victim's computer. Since the monetization part
of these operations often involves financial transactions with legitimate
institutions, researchers have identified the monetization of botnets as one
of the weak links of cybercriminal operations, and as a good point of
intervention for law enforcement~\cite{levchenko2011click}.
A second line of research focused on understanding the command and control
infrastructure used by
botnets~\cite{Nunnery:10:waledac,stone:09:torpig,cho:10:mega-d}.
These systems typically aim to reverse engineer the C\&C protocol with the goal
of infiltrating the botnet and collecting important information about the
cybercriminal operation~\cite{kanich:08:spamalytics} or developing systems to
detect and block such communication in the wild~\cite{Gu:2008:BCA}.
A third line of research focused on 
understanding the \emph{modus operandi} of cybercriminals using botnets, and what
makes their operations successful~\cite{iedemska2014tricks,StoneGross:11:leet}.
The focus of such research is to identify possible weak points in the workflow
followed by cybercriminals, and use such weak points for botnet mitigation. As
an example of such research, Stringhini et al.~\cite{stringhini2012babel}
discovered that spammers routinely clean up their email lists from non-existing
addresses by having their bots report back the error codes that they received
while sending emails. As a possible mitigation, they proposed that email servers 
send false replies to detected bots, forcing the botmaster to remove existing
addresses from their email lists, and reducing the amount of spam that such
servers end up receiving.

In this paper, we bring the understanding that we as researchers have of
botnet operations even further. We analyze the source code of the command and
control infrastructure of the Cutwail
botnet~\cite{StoneGross:11:leet,decker:09:cutwail}, which was one of the world's
largest spamming botnets between 2007 and 2012. This source code was
obtained as part of a takedown operation that involved academics, Internet
service providers, and law enforcement in late 2010. Having access to the source
code of the C\&C infrastructure provides us with a complete view on the logic
behind the command and control communication of a botnet, which so far could
have only been inferred by researchers from
observation~\cite{Nunnery:10:waledac,cho:10:mega-d}. This allowed us to identify
bottlenecks and vulnerabilities in the workflow required for C\&C communication, which could be used
by researchers and practitioners to cripple the effectiveness of the botnet.

For our experiments, we developed a stub bot
implementation to connect to the C\&C server, similar to what done by
researchers in the past~\cite{Caballero2011,stock:09:walowdac}. We then set up a
network of bots connecting to a C\&C server under our control, and performed a
number of attacks ran by the bots against their controller.
We show that misbehaving bots have the capability to enumerate the number of
bots currently registered with the C\&C server, impersonate an existing bot and
reporting false information to the C\&C server, and make the control server
inaccessible by mounting a distributed denial of service (DDoS) attack.
Interestingly, we show that 2,000 bots are enough to completely overwhelm the
C\&C channel and make the botnet non operational --- such number is much smaller
than the number of bots that C\&C servers can deal with in the
wild~\cite{iedemska2014tricks,StoneGross:11:leet,stone:09:torpig}.

The insights presented in this paper can help researchers and practitioners
develop better techniques to mitigate and cripple botnets. Although the analysis
was performed on a single botnet, many of our findings are generic and are due
to the workflow of command and control communication in general, rather than on
implementation problems. The attacks that are demonstrated in this paper could
be used to solve the problem of reliably enumerating the size of
botnets~\cite{fabian2007my}, deceiving botmasters by making them believe that
their bots are performing worse than they are, or that they have been
blacklisted, or helping practitioners and law enforcement deploy fake bots to
dilute the communication capability of the botnet and making it unusable for
the cybercriminal. 

In summary, this paper makes the following contributions:
\begin{enumerate}
 \item We present an analysis of the command and control infrastructure of
   Cutwail, a large spamming botnet. As part of this analysis, we provide a
   detailed description of the workflow and the logic behind the C\&C
   communication of spamming botnets. 
 \item We develop a number of attacks against the command and control logic of
   Cutwail. We set up a fake botnet in a restricted environment and demonstrate
   the feasibility and effectiveness of our attacks in gaining information about
   the botnet itself and crippling its operations.
 \item We discuss how our results generalize to other botnets different from
   Cutwail, and how similar techniques to the ones presented in this paper could
   be used by law enforcement and practitioners to take down botnets.
\end{enumerate}

\section{Background: the Cutwail botnet}

In this section, we provide an overview of the key components of the Cutwail botnet
and its propagation mechanism.

\subsection{Generic C\&C operations}
\label{subsec:operations}
To achieve its goals, a spamming botnet C\&C needs to perform three operations:

\begin{itemize}
\item Sending instructions to bots: the bots need to receive instructions
from the C\&C server in order to determine which emails to send, and to whom.
\item Communicating with bots in general: the C\&C server needs to manage its bots (e.g., keeping track of active bots) by communicating with them periodically.
\item Receiving reports from bots: the botmaster needs to receive spamming
statistics from the bots to measure the performance of the botnet, 
and to tune the botnet operation to make it more effective~\cite{iedemska2014tricks}.
\end{itemize}

Regardless of the botnet's implementation and topology,
the three operations described above are generic operations
of any spamming botnet.
Intuitively, the botmaster must be able to reach his bots 
to communicate and send instructions, and receiving reports from the bots 
for the outcome of their spamming operations is important
to measure the performance of the botnet, and to tune its operations.

The factors that make spam campaigns successful presented by
Iedemska et al.~\cite{iedemska2014tricks}, 
state that experience is what matters most for a spammer. 
Botmasters have to housekeep their botnets well, and by manually 
tuning botnet parameters, one can dramatically increase the outcome 
of spamming campaigns. 
The attacks that we present in this paper can be used to tamper with
statistics about the infected machines and overall 
spam operations stored in the C\&C database.
Since botmasters rely on this information to tune their botnets, 
these attacks can be used to deceive the botmaster into reducing
the effectiveness of his own botnet by providing false information.

\subsection{Cutwail botnet structure}

\begin{figure*}[!ht]
    \centering
    \includegraphics[width=10cm]{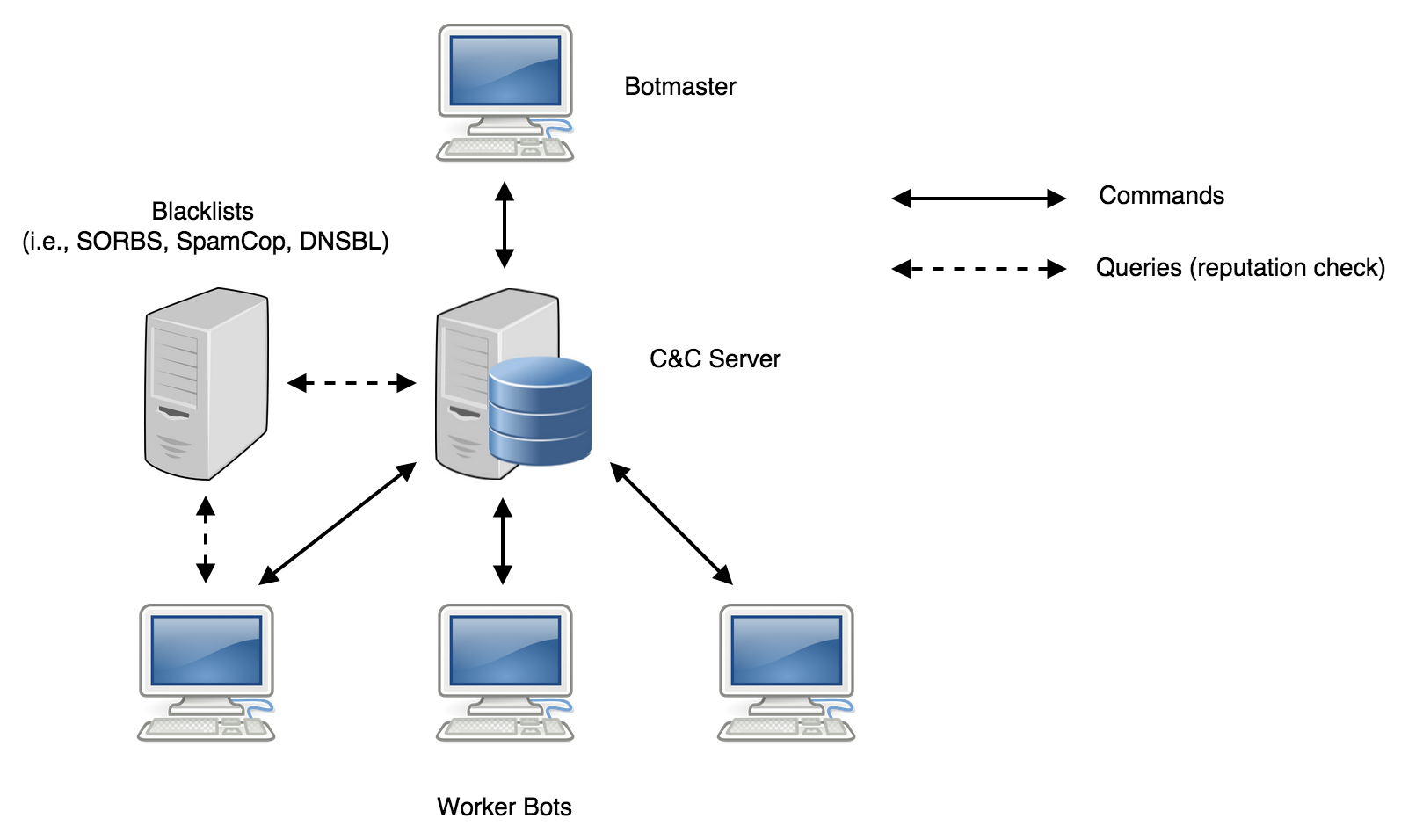}
    \caption{Schematic overview of the Cutwail botnet hierarchy.}
    \label{fig:hierarchy}
\end{figure*}

Cutwail has a fairly simple structure consisting of three different layers (as
shown in Figure \ref{fig:hierarchy}).
Firstly, the botmaster/spammer configures a spam campaign on the C\&C server.
Then, the bots connect directly to the C\&C server and receive instructions about 
emails they should send. After the co-opted bots have accomplished their task,
they report back spamming statistics (e.g., successful delivery, blacklisted 
by domain, etc.) to the C\&C server.

\subsubsection{Encrypted communication protocol.}
The original Cutwail botnet emerged in 2007, and has evolved 
in sophistication using simple HTTP request to a proprietary, 
encrypted protocol~\cite{StoneGross:11:leet}. 
The encrypted protocol is implemented using a block cipher in 
electronic codebook (ECB) mode. More details of the implementation 
of the protocol is available in~\cite{decker:09:cutwail}.

\subsubsection{Cutwail installation and infection process.}
A typical Cutwail infection occurs when a compromised machine executes a
so-called "loader" called Pushdo. Examples of infection vector include
drive-by download, or an attachment in a spam email. Pushdo behaves as 
an installation framework for downloading and executing various malware
components including rootkits that hide the presence of the malware in
the infected machine, and the Cutwail engine. After executing the Cutwail
engine, the Cutwail bot attempts to contact a command and control server
in order to receive serveral critical pieces of information to begin
a spam campaign. Specifically, the C\&C server provides the bot with
the actual spam content through "spam templates". More details in the
infection process and the technical aspects of the operation of the 
Pushdo loader are available in different studies
~\cite{StoneGross:11:leet,decker:09:cutwail}.

\subsubsection{Spam contents.}

The contents of the spam template include
(i) a list of target email addresses (also known as \textit{bases}) 
where a spam will be delivered.
(ii) a dictionary consisting of 71,377 entries for generating random
sender/recipient names and domains.
(iii) a configuration file containing details that control the spam engine's
behaviour (e.g., timing intervals, error handling, etc.).
The content of the email messages sent by Cutwail included pornography, 
online pharmacies, phishing, money mule recruitment and malware. 
The malware (e.g., the Zeus banking Trojan) is typically distributed 
by persuading a user to open an attachment in the form of greeting card, 
resume, invitation, mail delivery failure, and a receipt of recent purchase. 
In addition, many of the emails contained links to malicious websites that 
attempted to install malware on a victims system through drive-by-download 
attacks.

\subsubsection{Blacklisting.}

One of the most important aspects of a spam campaign is the ability 
to pass through both IP-based blacklists and content-based filters. 
Bots that are not blacklisted are the most valuable since they 
increase the chance of successfully delivering spam. 
Each Cutwail bot periodically queries several blacklists 
(i.e., SORBS, SpamCop, DNSBL), in order to determine its reputation 
(as shown in Figure \ref{fig:hierarchy}). This information is reported 
back to the C\&C server and recorded. The C\&C server also queries 
the blacklists periodically to determine the reputation of bots 
currently registered in its database.

In order to evade detection by content-based filters, a tool called 
\textit{macros} can be used to instruct each bot to dynamically generate 
unique content for each email by modifying fields such as sender address, 
email subject line, and body based on the spam template. 
Also, each Cutwail C\&C server runs a local instance of SpamAssassin, 
a free open source email spam filter based on content-matching rules. 
Once an email template has been generated, it is passed through 
SpamAssassin and tweaked until it successfully evades detection.

\subsubsection{Infiltrating Cutwail.}

Previous work on gaining insights into the operation of botnets 
via \textit{infiltration} (running clone bots, the so-called 
"milkers" in controlled environments) is available from
~\cite{Cho:10:II,Kreibich:2008:SCT,Kreibich:2009:SIL,stock:09:walowdac,Caballero2011}. 
Such work has primarily aimed at monitoring the instructions issued 
to bots in order to investigate how botmasters employ their botnets. 
In this paper, we bring forward the idea of employing milkers 
as a tool not only to monitor the Cutwail C\&C operations, but also 
to explore vulnerabilities in the C\&C workflow and logic to develop 
attacks against them.

\section{Analysis of the Cutwail C\&C software}

In this section, we present an analysis of the command and control
infrastructure of Cutwail. As part of this analysis, we provide a
detailed description of the workflow and the logic behind the C\&C
communication. We obtained the source code of the botnet by collaborating 
with Internet Service Providers and law enforcement during a takedown operation in 2010.

\subsection{Installation process}

The developer of Cutwail provides a shell script to assist the installation
of the Cutwail C\&C software. The software can be installed on a server
running either Linux or FreeBSD operating system.The installation script
first downloads libraries required to compile the program code, initialises
a MySQL database, and configures the SpamAssassin service. Then, it configures
the Makefile that generates five binary executable files, which are installed
under the \textit{/usr/local/psyche} directory.

The database consists of 34 tables that stores information about
bots and information required to operate spam campaigns. The \textit{bot}
table stores the bot identity number (BID), IP address, timestamps 
(e.g., last seen, born date), and spamming statistics for each 
registered bot. There are 17 status codes for reporting the delivery
result of a spam email, including SENT (status code: 1), NO\_USER (2),
BLACKLISTED (5), NO\_MX (8), SMTP\_TIMEOUT (11), and NO\_HOSTNAME (17).
The \textit{botstatus} table contains general information about the botnet, 
e.g., the number of bots currently online. 
The \textit{base} table has records that reference files containing
target emails addresses that are used by bulk operations.
The \textit{header}, \textit{message}, \textit{mailfrom}, and 
\textit{macros} tables contain information used to generate a spam template,
which is sent to the bots, and instructions for dynamically generating 
a unique spam content based on the template.

\subsection{Command and control program}

The main executable file responsible for running the spam operation is
called \textit{spcntrl} (spam control). Also an executable called 
\textit{spsupport} is run to support the spam operation by, for example, 
querying the IP-based blacklists in order to determine the reputation of 
bots registered in the database. Every time the \textit{spcntrl} program is
executed, it immediately computes and compares the hash code of the host's 
network interface configuration to the one generated during the installation process, and terminates if they do not match.
This is a mechanism for preventing security analysts from debugging the
program that has been moved onto a different environment.

After the program has successfully started, it loads what are called
\textit{common configurations} from the database. These are
general configurations that control the spam engine's behaviour,
which are independent from the configuration of each bulk operation.
Common configurations define constants such as the IP address of the 
C\&C server, the current version of the C\&C software, timing intervals, 
and the maximum number of bots the server can control. After loading 
these configurations, the control program creates three threads for 
managing bots, bulk operations, and TCP connections from port number 43,242.

\begin{figure*}[!ht]
    \begin{center}
    \includegraphics[width=12cm]{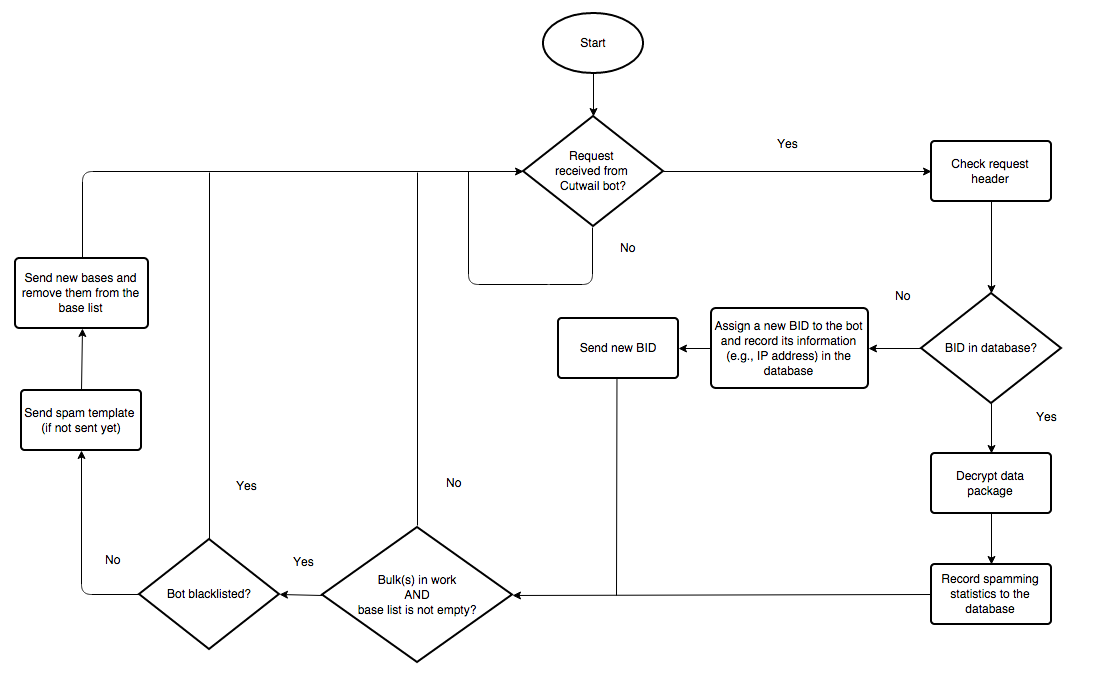} 
    \caption{Operation of the bot management thread.}
    \label{fig:botthread}
    \end{center}
\end{figure*}

Figure \ref{fig:botthread} shows a flowchart of the operation of the bot 
management thread.
Firstly, the C\&C server waits for a bot to establish a TCP connection and 
sends a valid request to the server.
After receiving the request, the server processes the header field, 
which contains the bot identifier number (BID) of the bot. 
If the BID is zero, the bot is identified as "new" and the server will 
assign a new BID to the bot and also record its information 
(e.g., BID, IP address, and timestamps) in the database. Otherwise, the 
server will expect an encrypted spamming report from the bot, which is 
decrypted and recorded in the database.

The C\&C server runs an Apache web server that hosts a web interface, which 
allows the botmaster/spammer to configure and manage bulk operations (or
spamming operations) from a browser. 
If there are any bulk operations currently in the 
"working" state, the server will send the corresponding spam template 
to the bot (if it has not been sent before). Also it will distribute a 
portion of the email database list if it is not empty, and the target email 
addresses (also known as bases) that are distributed to the bots are removed 
from the list. If the bot is identified as blacklisted, the server will not send 
any spam templates or bases to that bot.

The bot management thread has no mechanism for verifying the integrity of
the BID field in the server request header other than using it to determine
if a bot with that BID is currently registered in the database. 
This makes the command and control logic vulnerable to various exploits 
that are described later in this paper.

\subsection{Encrypted communication protocol}
\label{subsec:structure}

As previously described, Cutwail encrypts its communication using a
block cipher in ECB mode with an encryption key 29 characters long:
"Poshel-ka ti na hui drug aver"~\cite{decker:09:cutwail}.
After conducting a white box analysis by studying 13,904 lines of uncommented 
C source code and debugging the command and control program, 
we have gained an understanding in what each field in the server request 
mean, how they are processed, and how the server response is generated.

\begin{figure*}[!ht]
    \centering
    \includegraphics[width=12cm]{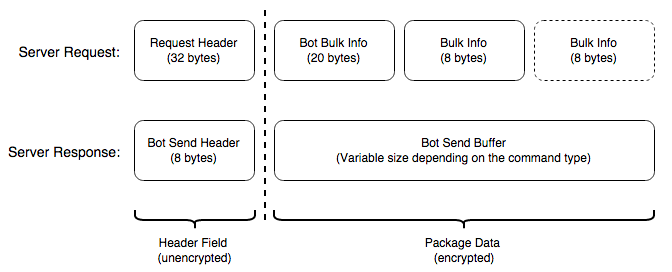}
    \caption{2-field type messages of server request and response.}
    \label{fig:package}
\end{figure*}

Figure \ref{fig:package} shows the dissection of the 2-field type messages
sent and received from the server. The server request consists of an 
unencrypted request header, followed by an encrypted data package
consisting of zero to one \textit{bot bulk info} structure, followed by 
zero or more \textit{bulk info} structures. The "size" field 
in the request header defines the size of the encrypted data package 
that follow (in bytes), which is used by the decryption function. 
Also, the header contains fields such as the BID, local IP address, 
Windows version, common configuration version, 
and the version of the bot (or the Pushdo loader). 
The \textit{bot bulk info} structure contains general information about 
the bulk operation that is assigned to the bot (e.g., the bulk ID, and the
spam template version number), and its "logsize" field defines the number of 
\textit{bulk info} structures, i.e., the number of spam email reports
that follow (default maximum of 1500). The email ID number and the delivery 
status code (previously described, e.g., SENT, BLACKLISTED, NO\_HOSTNAME, etc.) 
of each spam email is stored individually in each of the 
\textit{bulk info} structures.

The server response simply consists of an unencrypted response header, 
which contains the command type and the size of the encrypted data package
that follow. There are nine command types including 
RC\_SLEEP, RC\_GETWORK, RC\_RESTART, RC\_UPDATE, RC\_BID, and RC\_TEMPLATE 
(RC stands for Response Command), which determine the content of the 
data package, e.g., a new BID of the bot, bases, or the spam template.

\section{Cutwail clone implementation}

The Cutwail clone implements the encryption/decryption algorithm described
in ~\cite{decker:09:cutwail}, and the protocol operation that is used 
to communicate with the C\&C server is described below.

\begin{enumerate}
\item Establish a TCP connection with the C\&C server on port 43,242.
\item Send a server request with the structure described in section
\ref{subsec:structure} (with the BID initially set to zero), and wait for
the server response.
\item Upon receiving a response, extract the first four bytes of the
(unencrypted) header, which correspond to the command type, and the
remaining four bytes in the header to see the size of the encrypted 
data package. Decrypt the encrypted data package if its size is greater 
than zero.
\item If the command type is: 

\textbf{RC\_BID.} Extract the BID value from the decrypted data 
and change the BID of the clone bot accordingly.

\textbf{Otherwise.} Record the command type  and the decrypted
data (e.g., RC\_TEMPLATE and the spam template data), and continue.

\item Return to step 2.
\end{enumerate}

The clone bot only implements the communication feature of the botnet 
and does not cause any harm by sending spam emails, etc.
The only difference between a real bot and the clone 
is the interpretation of the command type in step 4.

\subsection{SSH botnet}
\label{subsec:sshbotnet}

We have set up a SSH botnet, which in our experiment consists of 19 virtual 
machines each capable of running up to 1024 instances of the 
Cutwail clone. This gives us the capability of controlling up to 
19,456 Cutwail clones, which can be used mount a distributed denial of
service attack against the Cutwail command and control server by just
instructing each clone to speak the communication protocol described above. 
We are by no means limited to using more virtual machines
for the SSH botnet, and if necessary, the number of virtual machines can be 
increased to raise the total population of the clone bots.

\section{Attacks against the Cutwail botnet}

In the following section, we describe four attacks against the
command and control logic of Cutwail that can be used to gain
information about the botnet itself and cripple its operations.
Specifically, we aim to exploit vulnerabilities in the three 
generic C\&C operations of spamming botnets (described in section \ref{subsec:operations}) 
that we can discover to disrupt the operation of Cutwail.
By doing this, we aim to showcase attacks that can be used for 
mitigation and takedown purposes against any spamming botnet, 
regardless of its specific implementation.

Firstly, the C\&C operation of sending instructions to bots is exploited by
using the clone bots to continuously request the C\&C server for bases, 
thereby preventing those bases to be received by real bots, and eventually, 
exhausting the base list maintained by the control server.
Secondly, the C\&C operation of communicating with bots in general is exploited
by mounting a distributed denial of service attack to saturate the server 
with external communication requests and make it respond so slowly 
as to be rendered non operational. We show that it is possible to completely disrupt the Cutwail C\&C operation by using 2000 bots, which is a much smaller number than the number of bots typically controlled by C\&C servers in the wild.
Finally, the C\&C operation of receiving spamming reports from bots is exploited
by using the clone bot to report false spamming reports on behalf of an arbitrary 
bot currently registered by the C\&C server. 
We also describe an attack to enumerate the number of bots currently registered in the
server database. Although this attack does not exploit any of the three generic
C\&C operations, it is used as an auxiliary attack for reporting fake spamming
reports.

\subsection{Exhausting the base list}

This attack exploits the generic C\&C operation of sending instructions to bots.
Each bulk operation must reference a file containing a finite list of bases 
(i.e., target email addresses) that are loaded during the start up of the 
\textit{spcntrl} program. These bases are distributed to bots upon request, 
and bases that have been distributed are removed from the list. 
When the base list becomes empty, the bulk operation simply stops distributing 
spam templates and bases to bots, and waits for spamming statistics to be reported.
Therefore, it is possible to constantly request the C\&C server from the clone 
bot to receive bases until the list of bases become exhausted/empty. This will prevent 
real Cutwail bots from receiving spam templates and bases, which are required 
to perform their spamming operations.

\subsection{Distributed denial of service attack}

This attack exploits the generic C\&C operation of managing bots by 
communicating with them in general.
We use the SSH botnet described in section \ref{subsec:sshbotnet}, to mount
a distributed denial of service (DDoS) attack against the botnet control server.
The C\&C server has a limited bandwidth and is limited to the number of 
concurrent connections it can manage from bots. 
Like other DDoS attacks, it is difficult to distinguish legitimate traffic 
from real bots and those generated by bots under our control.
This makes this attack difficult to defend against, and it will overload
the server by saturating it with external communication requests and/or 
making it respond so slowly as to be rendered non operational. 

\subsection{Enumerating the number of bots registered}
\label{subsec:enumeration}

This attack is used as an auxiliary task, and ise needed to report fake spamming reports.
In addition, this attack could be used to solve the problem of 
reliably enumerating the size of botnets.
Previous research \cite{StoneGross:11:leet,stock:09:walowdac,Cho:10:II} 
underlined the difficulty of estimating the size of botnets, which makes this attack particularly useful.

The BID column in the \textit{bot} table in the C\&C database is an 
auto-incremented primary key. When a new bot contacts the control server, 
it is given the largest BID in the current table incremented by one.
As previously mentioned in our analysis, the control server has 
no mechanism for verifying the integrity of the BID field in the 
server request header except for checking whether a record exists 
for that identifier in the table.
Therefore, it is possible to impersonate an 
existing Cutwail bot by just spoofing the BID field in the request header. 
An interesting behaviour is observed when doing this.
If a record for the spoofed BID already exists in the table, the server 
replies with the RC\_BID command, followed by a data package containing 
the same BID in the request. 
On the other hand, if the record does not exist in the table or the BID is 
equal to zero, the server identifies the bot as "new" and replies with a new BID.

Based on these observations, it is possible to enumerate the number of bots 
registered in the database by going through the following steps:
\begin{enumerate}
\item Firstly, send a request to the server with the BID field set to zero in the header.
\item The server will reply with the largest BID value in the table, incremented by one. 
This value is used as the upper bound to the number of bots currently registered in the 
database.
\item Send a server request with a spoofed BID field for each BID between one and the 
upper bound (obtained in step 2) decremented by one.
\item For each server response, compare the BID contained in the response header 
to the one in the request header. If they are equal, increment the bot count by one; 
otherwise, the BID does not exist in the table.
\end{enumerate}
The value obtained in step 2 is used as the upper bound, since some 
BID records less than that value is not guaranteed to exist because 
an experienced botmaster will remove records of bots that are performing 
badly to increase the effectiveness of his spam campaign. This is why 
steps 3 and 4 are executed to account for the missing BID records.

\subsection{Reporting fake spamming reports}

This attack exploits the generic C\&C operation of receiving spamming reports from bots.
After identifying the bot records that exist in the database 
(from the auxiliary enumeration attack), we can manipulate the spamming statistics 
of an arbitrary bot that is currently registered in the database 
by spoofing its BID and sending \textit{bot bulk info} 
and \textit{bulk info} data containing false spamming reports. 
As explained in section \ref{subsec:structure}, the \textit{bulk info} 
structure has a \textit{status} field, which can be set to any email delivery 
status, e.g., BLACKLISTED (status code: 5), which will cause the bot appear to be 
blacklisted by the domain of the target email address, for example, 
\textit{gmail.com}.
By making the bots appear to be performing worse than they are, botmasters may 
be deceived into abandoning those bots in attempt to increase the effectiveness 
of the botnet.

Also, we have observed that whenever a Cutwail bot establishes a TCP connection 
with the control server, the IP address, and the "last seen" field in the bot database 
is updated for the specified BID record. This means that by impersonating a currently 
registered bot, we can overwrite its IP address with the IP address of the clone, 
and the "last seen" field with the current time. Assuming that the \textit{real} bot 
is not going to connect back anytime soon, we can deceive the botmaster by making 
old or inactive bots appear to be active. Additionally, the IP address of the clone can be
spoofed to the one that is known to be in the blacklist (e.g., DNSBL), thus it
is possible to blacklist all the bots that are currently registered in the database
by overwriting their IP addresses. This could cripple the spamming operation
of the botnet as the C\&C server avoids distributing work to blacklisted bots.

\section{Evaluation}

This section presents the feasibility and effectiveness of each of 
the attacks described in the previous section.

\subsection{Experimental setup}

We have been careful to design experiments that we believe are
ethical. The attacks were tested in a controlled
environment with our Cutwail C\&C server and clone bots
running on VMware virtual machines with the \textit{host-only}
network configuration. However, by simulating the communication
between the control server and bots over a virtual network
adapter, we may have simulated the communication channel with
a higher bandwidth compared with that of the Internet.

\subsection{Exhausing the base list}
During the attack, a clone bot queried the C\&C server for some bases used 
for the bulk operation. As previously explained, the bases that are distributed
to the bots are removed from the base list containing a finite number of bases.
As a result, the base list maintained by the C\&C server quickly became empty, 
and the server stopped sending spam templates and bases to all the bots 
in the botnet thereafter.
By exhausting the base list, the attack will essentially prevent real Cutwail bots 
from receiving information required to perform their spamming operation. 
By default, Cutwail distributes 1000 target email addresses to each bot,
which means that 1000 clone bots would deplete a list of 1 million email addresses.
The attack will therefore, reduce the number of spam emails that are sent to the 
bases distributed to the clone bots, and thus, reducing the effectiveness 
of the spam campaign.
This attack can also be conducted in conjunction with the DDoS attack 
to use multiple clones to speed up the process of exhausting the list of bases.

\subsection{Distributed denial of service attack}
\begin{figure*}
  \begin{subfigure}[b]{0.5\textwidth}
    \includegraphics[width=\textwidth]{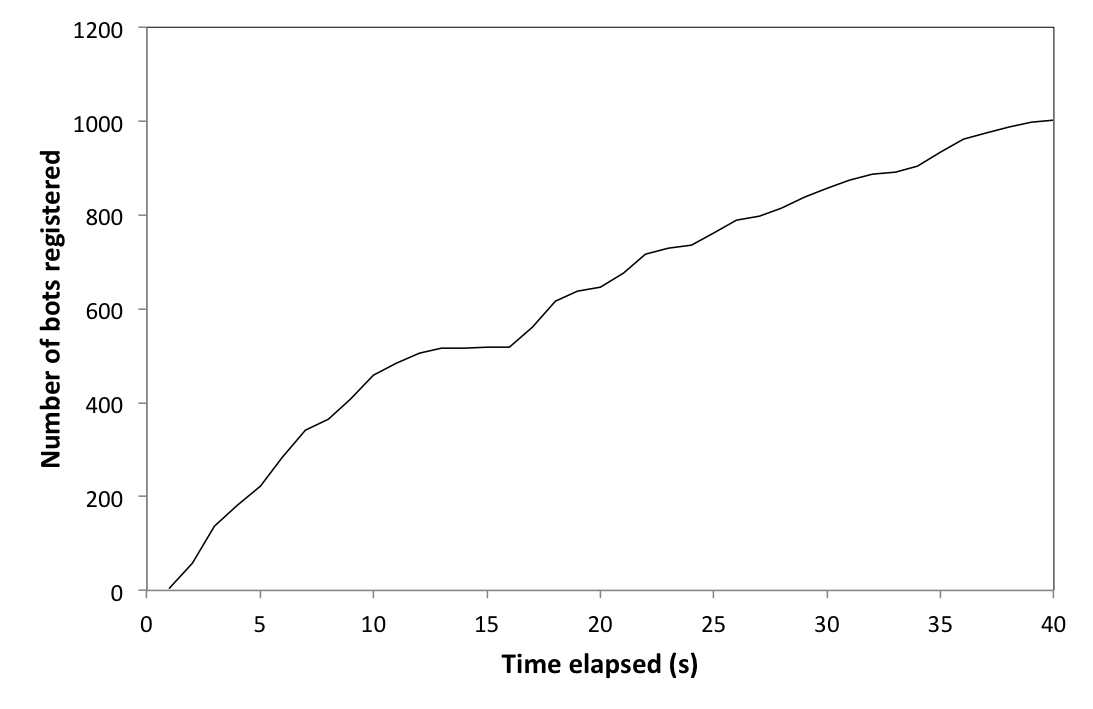}
    \caption{1000 clone bots}
    \label{fig:botregister1000}
  \end{subfigure}
  \begin{subfigure}[b]{0.5\textwidth}
    \includegraphics[width=\textwidth]{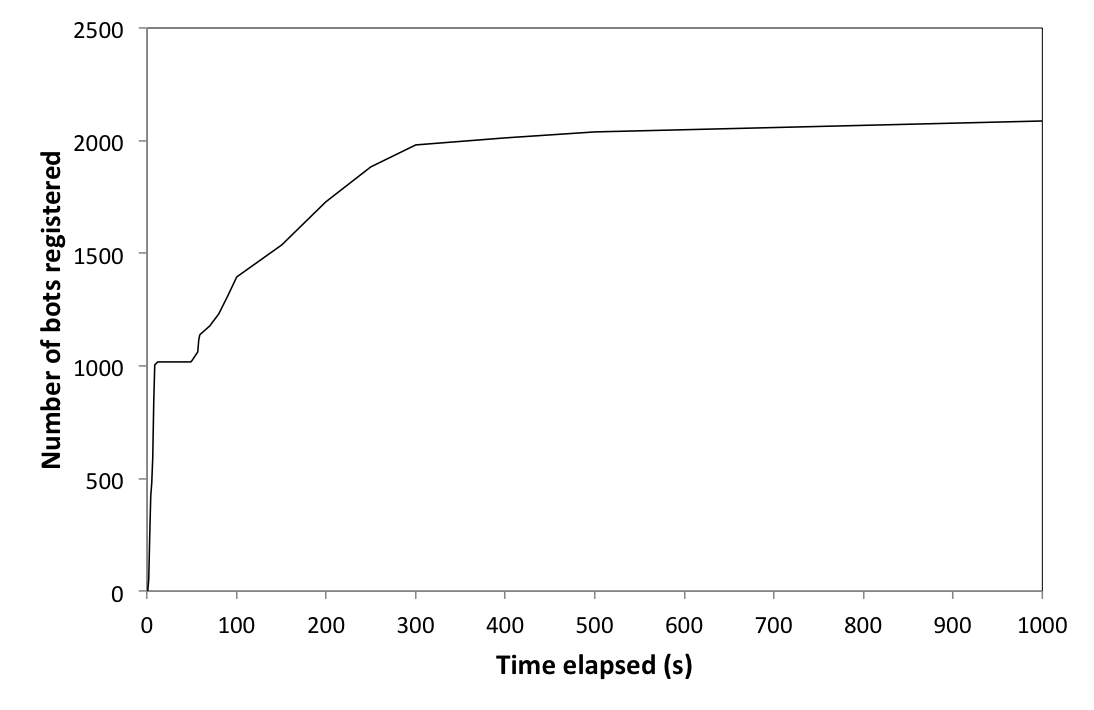}
    \caption{19456 clone bots}
    \label{fig:botregister19000}
  \end{subfigure}
  \caption{Number of bot registered against time during DDoS attack.}
  \label{fig:botregister}
\end{figure*}

\begin{figure}[!ht]
    \centering
    \includegraphics[width=7.5cm]{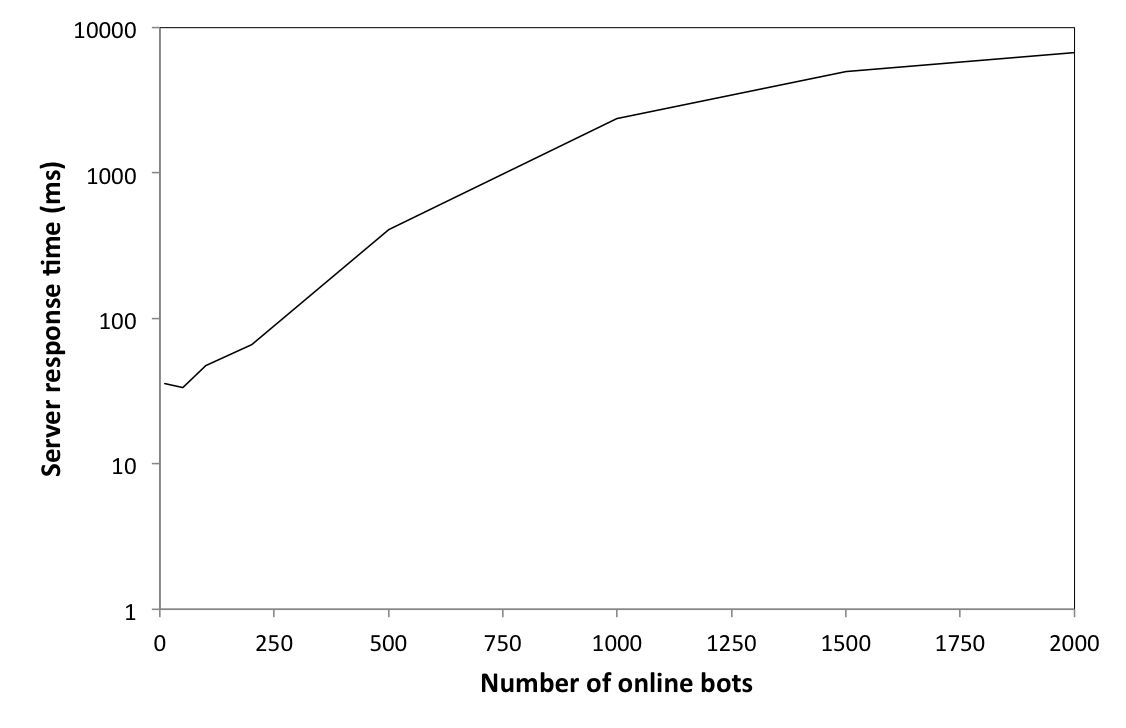}
    \caption{Server response time against number of online bots.}
    \label{fig:responsetime}
\end{figure}

To test the effect of the DDoS attack, the C\&C server is initialised with
a bulk operation with a base list containing 2048 entries. The rate at which
bots are registered (shown in Figure \ref{fig:botregister}) and the server 
response time against the number of online bots 
(shown in Figure \ref{fig:responsetime})
are measured while the server is attacked by 1000 bots or 19,456 bots controlled 
by the SSH botnet. During the course of the attack, each clone bot is instructed 
to overload the C\&C server by continually sending server requests.

In Figure \ref{fig:botregister1000}, we can see that the C\&C server manages to
register all 1000 clone bots in 40 seconds.
In Figure \ref{fig:botregister19000}, since 19,456 Cutwail clones are 
running, the C\&C server is expected to register around 19,456 bots. 
However, the number of bots that get successfully registered 
decreases dramatically once the server has registered above 2000 bots. 
The C\&C server at this point is overloaded with communication from the clone bots 
that are already registered, such that it cannot process requests from new bots. 
Therefore, the DDoS attack essentially prevents any new real Cutwail bots from 
even registering with the C\&C server.
It is interesting that only 2000 bots are enough to completely overwhelm the C\&C 
channel and make the botnet non operational - such a number is much smaller than 
the number of bots that C\&C servers can deal with in the wild. The only difference
is that the clone bots are instructed to flood the server with requests whereas 
real bots would only contact the server once they have accomplished their tasks.

Notice that the y-axis in Figure \ref{fig:responsetime} is using a logarithmic scale.
During the DDoS attack, as we increased the number of online (clone) bots, 
the server response time increased exponentially. The maximum number of bots used 
to record the response time was around 2000 due to the server overload.
This result shows that not only the DDoS attack prevents new bots from registering, 
but it also slows down the server response time exponentially to existing bots.
Furthermore, the botnet is expected to perform worse than the results presented,
since the communication channel on the Internet would have a narrower network bandwidth 
compared to that of the virtual network adapter used in the experiment.

\subsection{Enumerating the number of bots}
To simulate the situation where the Cutwail server has registered
some bots and is waiting for their response, the \textit{bot} table
in the database is initialised with dummy records with 100 as the
largest BID. The records for BIDs, 20 to 29 and 50 to 59, are
removed from the table to simulate poor performing 
bots deleted by the the botmaster.

\begin{figure}[!ht]
    \centering
    \includegraphics[width=8cm]{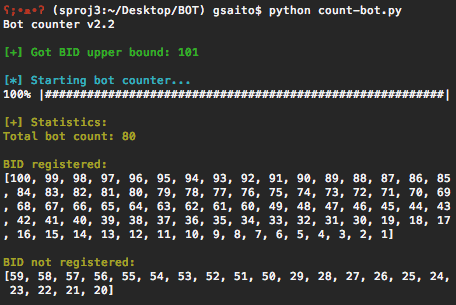}
    \caption{Result for enumerating number of bots currently registered.}
    \label{fig:botcount}
\end{figure}

The result of the attack for the experimental setup described above
is shown in Figure \ref{fig:botcount}. The attack successfully
enumerated the number of bots currently registered in the database.
However, previous research of
Cutwail by Stone-Gross et al.~\cite{StoneGross:11:leet} state that
while these BID values are unique, they do not appear to be an accurate
indicator of the total number of bots managed amongst different Cutwail 
C\&C servers. First, a Cutwail bot may connect to multiple C\&C servers 
over its lifetime, and thus, several C\&Cs may have their own identifier 
for a single bot possibly due to a bug in the malware.

Although we have devised a method for accurately counting the number of 
bots registered in a single C\&C server, it is not feasible to 
just add the results obtained from different servers to estimate 
the total population of bots managed amongst multiple servers.
There are other methods for estimating the total population of bots,
for example, by counting unique IP addresses of bots that connect 
to the C\&C server, but this will require eavesdropping the 
communication channel to the control server rather than attacking
the server itself.

\subsection{Reporting fake spamming reports}
The BIDs that exist in the database can be identified by running the
enumeration attack. In this experiment, the clone impersonates a "real" bot
by spoofing its BID in the server request header. The clone is
instructed to send a false spamming report (BLACKLISTED) 
on behalf of the real bot.
However, we also test the situation where the real bot is currently 
connected with the C\&C server in order to see how the C\&C server 
responds to connections from duplicate BIDs.

\begin{figure}[!ht]
    \centering
    \includegraphics[width=8cm]{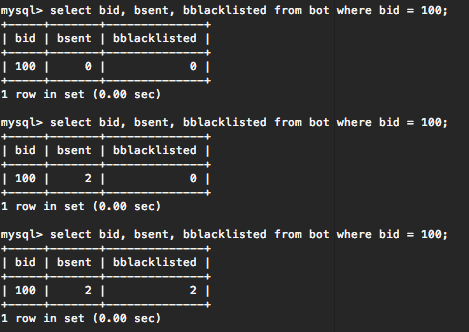}
    \caption{Result for reporting false spam delivery status.}
    \label{fig:falsestats}
\end{figure}

The result of reporting fake spamming reports is shown in 
Figure \ref{fig:falsestats}.
The real bot first reports the SENT status for two of the spam emails 
assigned. Then the clone bot reports the BLACKLISTED status, which is 
stored in the same BID record.
By debugging the command and control program, we have found that although
the real bot and the clone bot share the same BID, they are internally
treated as different bots by the control server and the BID value is only 
used to reference a row in the database to store the spamming statistics.
This logic could be exploited to manipulate the spamming statistics of an
arbitrary bot currently registered in the database regardless of whether
it is currently connected to the control server.

This attack could be used to deceive botmasters by making them believe 
that their bots are performing poorly. 
Previous research showed that successful spammers take this feedback into account, and stop using bots that are blacklisted or email addresses that are non existent~\cite{stringhini2012babel,iedemska2014tricks}.
This is an effective strategy for mitigating spamming operations of
botnets since it leads to a double bind for the botmaster/spammer: 
on one hand if the botmaster considers the feedback, he will remove a valid 
bot from his botnet. Effectively, this will reduce the size of the 
spamming botnet. On the other hand, if the botmaster does not consider 
the feedback, this reduces the effectiveness of his spam campaigns since 
the C\&C server will continue to use bots that are actually performing badly.

\section{Discussion}

The attacks presented in this paper may not be specific to the
Cutwail botnet since they exploit the generic C\&C operations of 
spamming botnets. The base list exhaustion attack, DDoS attack,
and reporting fake spamming reports all exploit the generic operations
of sending instructions to bots, communicating with bots in general,
and receiving reports from bots respectively.
Our findings have a number of implications as follows.

\subsubsection{Deceiving the botmaster.}
We have shown that it is possible to enumerate the number
of bots currently registered with the Cutwail C\&C server. 
Also by spoofing their BIDs, it is possible to impersonate
an arbitrary bot to report false information to the C\&C.
These attacks exploit the the generic operation of
bots reporting back the outcomes of their operations to the
C\&C server.
This gives us the capability of manipulating information that
botmasters use to tune their botnets, and they may be deceived
into abandoning bots that appear to be performing worse 
that they are or that they have been blacklisted, therefore 
reducing the size of the botnet.
Additionally, this will have an economic impact on the botmaster
as he will need to replenish his supply of bots, thinking that they are not suitable to the task anymore. 
This could lead him to buy new bots from the underground market or 
by using pay-per-install (PPI) services.

So far we have described the enumeration of the number of bots
as an auxiliary attack for impersonating an arbitrary bot in the botnet, 
however it can also be used to identify larger botnets in order to 
prioritize takedowns. Although the implementation of the enumeration attack 
described in section \ref{subsec:enumeration} is quite specific 
to the Cutwail botnet, similar techniques may be devised for other botnets by 
reverse engineering their command and control infrastructure.

\subsubsection{Crippling the effectiveness of the botnet.}
As we saw, we can exploit the generic operation of the C\&C server 
communicating with its bots in general by conducting a DDoS attack. 
We have shown that it is possible to saturate the server with
external communication requests as to prevent real Cutwail bots
from registering with the server. Also, we have observed that
the attack will increase the response time of the server
exponentially as we artificially increase the number of 
online (clone) bots. 
The most interesting result is that only
2000 clones were sufficient to overload the server, which is
much smaller than the number of bots the C\&C server can
manage in the wild (typically around 10,000 bots~\cite{iedemska2014tricks}).
Additionally, the generic operation of the C\&C server
sending instructions to its bots can be exploited by exhausting the
base list maintained by each bulk operation run by the C\&C server.
A consequence of this is that the C\&C server distributes most of 
the bases to the clone bots instead of the the real bots, 
hence the number of spam emails that will be sent to those 
email address can be reduced.

Since all of the attacks are implemented by just instructing the
clone bot to speak the communication protocol of the botnet, 
in order to defend against the attack the botmaster has to first solve 
the problem of distinguishing legitimate server requests with 
those generated by the clones.

Although the analysis was performed on a single botnet
(i.e., Cutwail), the insights presented in this paper can help 
law enforcements and practitioners develop better techniques 
to mitigate and cripple other spamming botnets since many of our finding 
are generic and are due to the workflow of command and control 
communication in general
(e.g., distributing bases, communicating with bots, and reporting spamming statistics), 
rather than on implementation problems.

\section{Related Work}

Computer security researchers have paid a considerable amount of attention to
the threats posed by botnets and their operations~\cite{Cooke:2005:ZRU}.
Existing research in the field of botnets mostly falls in two categories:
botnet analysis and botnet mitigation.

\noindent\textbf{Botnet Analysis.} At their beginning, botnets mostly used
Internet Relay Chat (IRC) as their C\&C channel.
In 2006, Abu Rajab et al. tracked 192 unique IRC botnets and gained precious insights
on how these networks operated~\cite{Rajab:2006:MAU}.
Once botnets moved away from IRC and started using proprietary C\&C protocols,
researchers started writing their own bots speaking such protocols and
infiltrated multiple
botnets~\cite{Cho:10:II,Kreibich:2008:SCT,Kreibich:2009:SIL,stock:09:walowdac,Caballero2011}.
Their analysis and results provided valuable information on how botnets are
operated. In this project, we implemented our own stub bot to speak Cutwail's
C\&C protocol. Unlike previous work, who had to reverse engineer the protocol
from observations or binary analysis, we were fortunate enough to have access
to the server-side code.
Chiang et al. studied the Rustock spambot and provided an analysis
of this spamming botnet, which was the most active between 2010
and 2011~\cite{Chiang:2007:CSR}. Decker et al. performed a similar analysis
based on the Pushdo trojan~\cite{decker:09:cutwail}.
Rossow et al. presented a taxonomy of peer-to-peer
botnets~\cite{rossow2013sok}.

Recently. Nadji et al. discussed how to perform effective botnet
takedowns~\cite{nadji2013beheading}. The issue of botnet takedowns is very
complex, and this paper could help practitioners and law enforcement in
devising possible techniques to perform effective ones.

Previous analysis of the Cutwail botnet was presented by Stone
Gross et al.~\cite{StoneGross:11:leet}, and it was based on the data collected
from multiple C\&C servers as a result of an attempted takedown.
Our work integrates the one conducted previously, offering a detailed view of
the C\&C protocol used by Cutwail as well as presenting some weak points in
the control workflow of the botnet, which could be used to cripple it and make
it less efficient. In addition, we show how our observations could be applied to
different botnets than Cutwail, because many of them tackle weaknesses in the workflow
of a botnet, rather than in its specific implementation.

\noindent\textbf{Botnet mitigation.} A number of projects dealt with
automatically reverse engineering the C\&C protocol used by botnets, by
performing dynamic analysis on the bot binaries. Such systems allowed
researchers to gain a better understanding of the C\&C protocols, and use that
knowledge to infiltrate multiple
botnets~\cite{Caballero:2009:DEA,Prospex09,Wondracek:08:ANPA,Polyglot07,Lin:08:APF,inference10}.
In this project, we analyzed the server-side code of the C\&C communication of a
large botnet. This allowed us to get a better understanding of the workflow
followed by botmasters in managing their bots, and identifying weak points in
such workflow.

Other work focused on identifying C\&C traffic in network data, and use this for
detection~\cite{Perdisci:2010:BCH,Ehrlich:2010:DSH,Gu:2008:BCA,Yen:2008:TAM,WurzingerBHGKK09}.
A problem that this type of projects faces is the increasing use of cryptography
by botmasters, which makes the creation of signatures difficult. In this
project, having access to the source code of the C\&C server allowed us to get a
deep knowledge of the cryptographic protocol used by Cutwail, without need to
reverse engineer it.

Other work looked at the activity performed by bots and used that for detection.
PRominent examples used the email spam activity of bots to this
purpose~\cite{Pitsillidis:10:BotnetJudo,Qian:2010:NLC,John:2009:SSB,Xie:2008:SB,Zhao:2009:BLS}.
The observations in this paper are more general, and could generalize to botnets
that are used for other purposes, such as performing DoS attacks.

Stringhini et al.~\cite{stringhini2012babel} proposed a system to provide false
information to botmasters and makes their operations less effective. Their
system works by having a mailserver send false information to a known bot (for
example a blacklisted one). In this paper, we bring forward this idea, and show
the feasibility of creating fake bots that impersonate existing ones, and
provide the botmaster with false information about them, forcing him to dropping
such bots and purchasing new ones. We believe that this type of strategy could
help greatly in the task of botnet mitigation.

\section{Conclusions}
We have presented an analysis of the command and control infrastructure
of Cutwail, one of the world's largest spamming botnet between 2007 and 2012.
Also we have developed a number of attacks against the command and control logic
of the server,
which were made possible by setting up a network of clone Cutwail bots in a controlled environment.
Our experiments show that misbehaving bots have the capability of
not only being able to extract information about spamming campaigns
operated by the botnet, but also to manipulate critical information
stored in the server database to deceive the botmaster, and to cripple 
the effectiveness of the botnet by overloading the control server.
Our inside view of the command and control infrastructure of Cutwail and the 
infiltration strategies developed offer new insights to law enforcement and 
practitioners to devise similar techniques to take down other botnets.

\bibliographystyle{acm}
\bibliography{biblio}
\end{document}